\documentclass[12pt]{article}
\usepackage{amssymb,amsmath}
\begin{document}

\title{\bf Power-laws from critical gravitational collapse: The mass
  distribution of subsolar objects}

\author{{\bf Matt Visser}\\
School of Mathematical and Computing Sciences, \\
Victoria University of Wellington, \\ 
P.O. Box 600, Wellington,  New Zealand\\[20pt]
{\bf Nicolas Yunes}\\
Center for Gravitational Physics and Geometry, \\
Center for Gravitational Wave Physics, \\
Department of Physics, \\
The Pennsylvania State University,\\
University Park, Pennsylvania 16802, USA}

\date{15 March 2004; \LaTeX-ed \today}

\maketitle

%
%
\def\d{{\mathrm{d}}}
\def\implies{\Rightarrow}
\def\be{\begin{equation}}
\def\bd{\begin{equation}}
\def\ba{\begin{eqnarray}}
\def\bea{\begin{eqnarray}}
\def\ee{\end{equation}}
\def\ed{\end{equation}}
\def\ea{\end{eqnarray}}
\def\eea{\end{eqnarray}}
\def\ie{{\emph{i.e.}}}
\def\eg{{\emph{e.g.}}}
\def\p{{\partial}}
\def\N{I\!\!N}

\begin{abstract}
  Critical gravitational collapse and self similarity are used to
  probe the mass distribution of subsolar objects. We demonstrate that
  at very low mass the distribution is given by a power law, with an
  exponent opposite in sign to that observed at high-mass regime. We
  further show that the value of this low-mass exponent is in
  principle calculable via dynamical systems theory applied to
  gravitational collapse. Qualitative agreement between numerical
  experiments and observational data is good.

  \bigskip

  \centerline{astro-ph/0403336; CGPG-03/10-6}
\end{abstract}

\clearpage

\section{Background}

At large mass, the initial mass function [IMF] describing the mass
distribution of stellar objects is characterised by a power law with
the Salpeter exponent $1.35$.  At small subsolar mass, we demonstrate
in a model-independent manner that there \emph{must} be a change in
this power law, and that the sign of the exponent must flip.  Direct
observation indicates that the IMF is certainly modified below
approximately $0.8 \; M_{\odot}$, and we confront theoretical
expectations and numerical simulations with the observational data.

Gravitational condensation, either Newtonian or general relativistic,
is characterised by the existence of critical exponents and power law
behaviour.  By linearising around any critical solution at the threshold
of collapse, the mass $M$ of the resulting condensed object is
related to any suitable control parameter $A$ in the initial data by
an equation of the form~\cite{Harada03,Harada01,Gundlach}
\be 
M \approx M_0 \; [A-A_{\mathrm{critical}}]^\delta;
\qquad \delta > 0.  
\ee 
Once a scaling law of this type is derived, straightforward
manipulations lead to a power law for the distribution of low-mass
objects
\be
P(M) \approx \frac{\cal A}{M_0} \; 
\left(\frac{M}{M_0}\right)^{(1/\delta)-1},
\ee
with an exponent that is calculable in terms of the mass-scaling
exponent. In this manner, we can explain the low-mass tail in the
Initial Mass Function [IMF] from first principles in terms of
dynamical systems theory in gravitational collapse. The technique
developed in this article cannot say anything about the high-mass tail
of the IMF, but that is a regime where there is reasonable theoretical
and observational agreement on the state of affairs. We shall
specifically concentrate on the functional form of the IMF for
subsolar masses.
 
To set the stage, recall that any gravitationally self-interacting
cloud of gas, either Newtonian or general relativistic, has a limited
number of long-term fates:
\begin{itemize}
\item The cloud can completely disperse to infinity.
  
\item Part of the cloud might condense, with the remainder dispersing
  to infinity.
  
\item The entire cloud might condense.
\end{itemize}
The condensed object could, for instance, be a solid planet, a fluid
star, or a black hole. The set of all initial data that lead to any
one of these fates can be thought of as an infinite-dimensional phase
space, containing infinite-dimensional basins of attraction for each
final fate. Since there are three possible final fates for a cloud of
gas, there will be three basins of attraction: the collapse basin,
where its attractor leads to complete collapse; the dispersal basin,
for which the final fate is an asymptotically flat Minkowski
spacetime; and an intermediate collapse basin, where ultimately part
of the cloud collapses and the rest disperses to infinity. These
basins will be separated from each other by boundaries of co-dimension
one, or separatrices, that form the so-called critical surfaces. In
this manner, it is clear that the critical surfaces contain all
critical initial data that separate two basins of attraction.  An
example of critical initial data, {\it{i.e.}} a point on the critical
surface, would be the Jeans mass, or Jeans energy.  Another important
point on this surface will be an intermediate attractor in phase
space, and it will be referred to as the critical solution. This
critical solution will have important properties, such as
self-similarity or scale-invariance.  For a more complete and detailed
analysis refer to~\cite{Gundlach}.

Applying dynamical systems theory to the region of phase space close
to the collapse-dispersal separatrix leads generically to the
prediction of power-law behaviour for the mass of the resulting
condensed object.  In order to make this point more explicit, let us
consider some set of initial data parameterised by the control
parameter $A$. Let us also assume that for $A < A_\mathrm{critical}$
the cloud completely disperses, while for $A > A_\mathrm{critical}$ at
least part of the cloud condenses. In other words, if $A$ lies inside
of the intermediate collapse basin, the solution to the field
equations will be equivalent to finding an integral curve in phase
space from $A$ to the final attractor of this basin. Similarly, if $A$
lies inside of the dispersal basin, then the integral curve will start
at $A$ but end at the final attractor of dispersal.  Then, under the
mild technical assumption of the existence of at least one critical
collapse solution on the critical surface with exactly one unstable
mode~\cite{Harada03,Harada01,Gundlach}, the condensed mass will be
given by
\be 
\label{scaling}
M \approx M_0 \; [A-A_\mathrm{critical}]^\delta, 
\ee 
provided that the initial data is chosen reasonably close to the
critical surface, {\it{i.e.}} $A \approx A_\mathrm{critical}$.

The physically interesting quantity is the exponent $\delta$, which
arises naturally as the fractional power-series exponent of a
generalised Frobenius expansion for linear perturbations around the
critical solution~\cite{Yunes}.  The order parameter $A$, the critical
initial data $A_\mathrm{critical}$, and the constant of
proportionality $M_0$ can be changed at will by reparametrising the
initial data set. In contrast, the exponent $\delta$ is physically
significant, depending only on the equation of state and the
condensation mechanism. Observe that, by construction, we must have
$\delta >0$, since $\delta<0$ would imply an abrupt transition between
no condensation and complete condensation of the cloud. Even
$\delta=0$ is problematic, since this corresponds to an abrupt
transition from no condensation to a finite condensate mass. It is
only for $\delta>0$ that as we fine-tune the control parameter $A$ we
get the physically reasonable situation of no condensation connected
smoothly to a low mass condensate for $A > A_\mathrm{critical}$.

Behaviour of this type has now been seen in a number of seemingly
disparate situations. In Newtonian gravity coupled to a gas cloud with
some specified equation of state, such as an isothermal one, it is
possible to observe the same scaling behaviour of the mass. First, one
searches for solutions describing critical collapse and then
linearises around these critical collapse solutions to find
$\delta$~\cite{Harada03,Harada01}. The Newtonian isothermal collapse
case is of particular relevance in astrophysics, since it is a good
description for cold molecular gas in the interstellar medium, where
the cooling time is much shorter than the dynamical time.  In general
relativity, the special case where the condensed object is a black
hole is known as Choptuik scaling~\cite{Gundlach,Choptuik}. This
phenomenon has now been analysed not just for gas clouds but also for
several other forms of matter. In particular, the analysis in
\cite{Choptuik98,koikehara} showed that for an adiabatic perfect fluid
with adiabatic index in the domain $\gamma \in \left(1,1.89\right)$,
where $p=(\gamma-1)\rho c^2$, the critical exponent varies over the
range $\delta \in \left(0.106,0.817\right)$, clearly demonstrating the
dependence of this exponent on the equation of state.  Several key
results are summarised in Table I.

\bigskip

\begin{table}[htb]
\begin{center}
Critical exponents determined by numerical experiment.

\medskip
\begin{tabular}{||l|l||l|l||}
\hline
\hline
System & Critical Point & Exponent $\delta$ & $1/\delta$ \\
\hline\hline
Newtonian isothermal & Hunter A &  0.10567  & 9.4637\\
\hline
GR dust: $p=0$ & Evans--Coleman & 0.10567 &  9.4637   \\
GR radiation: $p=\frac{1}{3}\rho c^2$ & Evans--Coleman & 0.3558019 & 2.810553 \\
GR semi-stiff: $p=\frac{4}{5}\rho c^2$ & Evans--Coleman & 0.73 & 1.37 \\
GR stiff: $p=\rho c^2$ &  Evans--Coleman & 0.96 &   1.04\\
\hline\hline
\end{tabular}
\\[10pt]
{Table I: Key known values of critical exponents in various systems.
  \\ See
  references~\cite{Harada03,Harada01,Gundlach,Choptuik98,koikehara}
  and references therein.}
\end{center}
\end{table}

\section{From critical collapse to IMF}

Extending this analysis further, suppose a number of Newtonian
systems, with initial data depending on some control parameter $A$,
evolve dynamically.  Let the distribution of initial control
parameters be given by the probability distribution function $P_a(A)$.
We can then determine the probability $P(M)\propto \mathrm{d}N /
\mathrm{d}M $ of producing low-mass condensed objects by calculating
\be
P(M) \; \d M = P_a(A) \; {\frac{\d A}{\d M}} \; \d M. 
\ee
We can use Eq.~\ref{scaling}, to rewrite the probability as
\be
P(M) \; \d M \approx {\frac{1}{\delta}} \; {\frac{P_a(A_\mathrm{critical})}{M_0}} \;  
\left(\frac{M}{M_0}\right)^{(1/\delta)-1} \; {\d M}.
\ee
Therefore, regardless of what the probability distribution $P_a(A)$
looks like, as long as it is smooth near $A_\mathrm{critical}$, we
expect for low mass objects a power law distribution in masses:
\be
P(M\ll M_0) \approx 
{\frac{\cal A}{M_0}} \; \left({\frac{M}{M_0}}\right)^{(1/\delta)-1}.
\ee
Observe that this analysis holds only for small masses, since we have
assumed that the control parameter $A$ is near the critical surface.
This behaviour is structurally similar to the observed high-mass IMF,
\be 
\zeta(M) = \int P(M) \; \d M,
\ee
given by a probability function with a power law of the form
\be 
P(M\gg M_0) \approx {\frac{\cal B}{M_0}} \; 
\left(\frac{M}{M_0}\right)^{-m-1},  
\ee
where observation favours the Salpeter exponent $m\approx 1.35$. The
major difference is that at low mass the sign of the exponent changes,
which is necessary on two counts: in order that the probability
function be integrable, and that the exponent $\delta$ be even in
principle calculable within the current scenario.  A simple toy model
that exhibits both forms of asymptotic behaviour is
\be
P(M) =  {\frac{n \; m}{n+m}} \; {1\over M_0} \; \left\{
\left(\frac{M}{M_0}\right)^{+n-1}  \Theta(M_0-M)
+
\left(\frac{M}{M_0}\right)^{-m-1}  \Theta(M-M_0)
\right\},
\ee
where both $n$ and $m$ are positive. 

\section{Observational situation}

In contrast to these theoretical considerations, direct astrophysical
observation leads to several models for $P(M)$ that are piecewise
power laws (Table II), and to several isolated data points at low mass
(Table III). The three standard IMFs are those of
Salpeter~\cite{Salpeter}, Miller--Scalo~\cite{Miller-Scalo}, and
Scalo~\cite{Scalo}, with a more recent version due to
Kroupa~\cite{Kroupa}. Relatively few of the ranges in Table II
correspond to a positive $\delta$. For low mass condensates, Scalo
gives $m = -1/\delta= -2.60$ so that $\delta =0.385$, while Kroupa
gives $m = -1/\delta \in ( -1.4, 0.0) $ so that $\delta \in ( 0.71,
\infty)$. All the other parts of the standard IMFs correspond to the
high mass region where the number density is decreasing with
increasing mass.

\begin{table}[htb]
\begin{center}
Multi-scale observational IMFs.

\medskip
\begin{tabular}{||l|c|c||c||}
\hline
\hline
IMF: $P(M) = ({\cal A}/M_0) \; (M/M_0)^{-m-1}$ &
 $M_1/M_\odot$ & $M_2/M_\odot$ & Exponent $m$\\
\hline\hline
Salpeter~\cite{Salpeter} & 0.10  & 125 & 1.35\\
\hline\hline
Miller--Scalo~\cite{Miller-Scalo} & 0.10& 1.00 & 0.25\\
 & 1.00& 2.00& 1.00\\
 & 2.00& 10.0& 1.30\\
 & 10.0& 125& 2.30\\
\hline\hline
Scalo~\cite{Scalo} & 0.10& 0.18 & $-2.60$\\
 & 0.18 & 0.42 & 0.01\\
 & 0.42& 0.62 & 1.75\\
 & 0.62& 1.18 & 1.08\\
 & 1.18 &3.50 & 2.50\\
 & 3.50& 125& 1.63\\
\hline\hline
Kroupa~\cite{Kroupa} & 0.01 & 0.08 & $-0.7\pm 0.7$ \\
 & 0.08 & 0.50 & $+0.3\pm0.5$ \\
 & 0.50 & $\infty$ & $1.3\pm0.3$ \\
\hline\hline
\end{tabular}
\\[10pt]
{Table II: Observationally derived piecewise power-law $P(M)$.}
\end{center}
\end{table}

Those IMFs obtained using observations which focused on the substellar
regime are summarised in Table III. These observations indicate broad
observational agreement as to the sign of the low-mass exponent, and a
preponderance of evidence pointing to a clustering of the exponent at
$m\approx -0.5$, {\it{i.e.}} $n\approx +0.5$ and $\delta \approx +2$.
These low-mass exponents are converted into critical exponents in
Table IV.  By comparing the theoretical results in Table I with the
observational results in Table IV, we can see that while there is
broad agreement between observation and theory regarding the sign of
the exponent, \emph{quantitative} agreement is more problematic.

\begin{table}[htb]
\begin{center}
Low-mass observational IMF.

\medskip
\begin{tabular}{||l|c|c||c||}
\hline\hline
IMF: $P(M) = ({\cal A}/M_0) \; (M/M_0)^{-m-1}$ &
 $M_1/M_\odot$ & $M_2/M_\odot$ & Exponent $m$\\
\hline\hline
Barrado y Navascues et al~\cite{Barrado00} &0.2& 0.8 & $-0.2$\\
Barrado y Navascues et al~\cite{Barrado} &0.035& 0.3 & $-0.4$\\
Bouvier et al~\cite{Bouvier}  & 0.03 & 0.48 & $-0.4$\\
Mart\'\i{}n et al~\cite{Martin} &0.02&0.1  & $-0.47$\\
Bouvier et al~\cite{Bouvier0209}  & 0.072 & 0.4 & $-0.5$ \\
Luhman \& Rieke~\cite{Luhman} & 0.02 & 0.1  & $-0.5$ \\
Najita et al~\cite{Najita} & 0.015 & 0.7 & $-0.5$\\
Rice et al~\cite{Rice} 
& $10^{-5}$ & $10^{-3}$ & $\approx -1$\\
\hline
Tej et al~\cite{Tej} & 0.01 & 0.50 & $-0.2\pm 0.2$\\
& 0.01 & 0.50 & $-0.5\pm 0.2$ \\
\hline\hline
\end{tabular}
\\[10pt]
{Table III: Observationally derived low-mass $P(M)$.}
\end{center}
\end{table}

\bigskip

\begin{table}[htb]
\begin{center}
Observed low-mass exponents.

\medskip
\begin{tabular}{||l|l||l|l||}
\hline
\hline
Source & Exponent $m$ & Exponent $1/\delta$ &  Exponent $\delta$ \\
\hline\hline
Scalo~\cite{Scalo} & $-2.60$ & 2.60 & 0.385 
\\
Kroupa~\cite{Kroupa} & $-1.4$ --- $0.0$ &$0.0$ --- $1.4$ & $0.71$ --- $\infty$
\\
\hline
Rice et al~\cite{Rice}  &$\approx -1$ & $\approx 1 $ & $\approx 1$
\\
Najita et al~\cite{Najita} & $-0.5$ & 0.5 & 2.0 
\\
Luhman \&  Rieke~\cite{Luhman} & $-0.5$ & 0.5 & 2.0 
\\
Bouvier et al~\cite{Bouvier0209}  &  $-0.5$ & $0.5$ & $2.0$
\\
Mart\'\i{}n et al~\cite{Martin} & $-0.47$ & $0.47$ & $2.16$
\\
Bouvier et al~\cite{Bouvier}  &  $-0.4$ & $0.4$ & $2.5$
\\
Barrado y Navascues~\cite{Barrado} & $-0.4$ & 0.4 & 2.5
\\
Barrado y Navascues~\cite{Barrado00} & $-0.2$ & 0.2 & 5.0
\\
\hline
Tej~\cite{Tej} & $-0.5$ & 0.5 & 2.0\\
& $-0.2$ & 0.2 & 5.0 
\\
\hline\hline
\end{tabular}
\\[10pt]
{Table IV: Observational estimates of the very low mass exponents.}
\end{center}
\end{table}

We must conclude that present day observational data is sufficiently
poor that the only rigorous inference one can draw is that the
exponent has changed sign at sufficiently low masses. Beyond that, it
would be desirable to contrast the exponent occurring in the subsolar
IMF with the exponent arising in a specific critical collapse process.
Unfortunately, neither observational data nor theory is currently well
enough developed to do so with any degree of reliability.  Some of the
numerical simulations give critical exponents that overlap with some
of the observations. For instance, the Scalo exponent is roughly
comparable with that arising from numerical simulations of collapse of
a relativistic radiation fluid, $p={\frac{1}{3}}\rho c^2$.  Part of
the range of Kroupa's IMF, {\it{i.e.}} $\delta\in(0.71,1)$, is
compatible with simulations of a relativistic adiabatic perfect fluid,
$p = k\;\rho c^2$ with $k\in({\frac{4}{5}},1)$. Finally, the IMF
exponent of Rice \emph{et al} is compatible with a numerical critical
solution corresponding to a relativistic stiff fluid, $p = \rho c^2$.
Those observations that cluster around $\delta=2$ are not compatible
with any \emph{known} critical collapse solution. This might indicate
either a problem with the observational data, or a more fundamental
lack of understanding regarding the physically relevant critical
collapse process.

For instance, a plausible explanation for step-wise changes in IMF
exponents is to consider the possibility that there are several
competing collapse processes with different critical solutions.  If
this is the case, all such solutions that have a single unstable mode
will contribute to the IMF, leading to a probability distribution of
the form
\be
P(M) \approx \sum_i {\frac{{\cal A}_i}{M_0}} \; 
\left(\frac{M}{M_0}\right)^{(1/\delta_i)-1}.
\ee
This leads to a ``kinked'' power law law where the largest of the
$\delta_i$ dominates at smallest masses.  Eventually, there will be a
switch-over to one of the other critical exponents at larger masses.
If this larger mass is still reasonably small, one could still
calculate using dynamical system theory.  In this manner, one may hope
to model the IMF all the way up to its peak. One can never, however,
obtain the high-mass decreasing tail from this sort of analysis.

\section{Conclusions}

Future work along these lines should be focused in two directions.
Observationally, improved data would be desirable to test the
hypothesis that the low-mass exponent $\delta$ is both positive and
universal.  Theoretically, it would be important to understand
\emph{quantitatively} why critical behaviour provides an accurate
representation of the IMF for $M \lesssim 0.8 ~M_{\odot}$.  It is
clear that as the final condensed mass increases, the initial data $A$
is pushed farther away from the critical surface, {\it{i.e.}} $A \neq
A_\mathrm{critical}$.  Although it is known that the linear
perturbation around the critical solution then loses validity, a
precise calculation of the region of convergence is still lacking.
Furthermore, since the formation of real-world gravitational
condensates is likely to involve rotating turbulent dust clouds, it
would be very useful to understand the influence of both angular
momentum and turbulence on the theoretically derived critical
exponents.

Our analysis confirms Larson's intuition that stellar formation at low
mass is related (and perhaps even dominated) by chaotic
dynamics~\cite{Larson}.  In particular, the analysis in terms of
dynamical systems theory can be viewed in terms of deterministic chaos
in gravitational collapse.  We do not, however, need to deal with
fractal structures since limit points and limit cycles seem to be
quite sufficient for generating power-law behaviour~\cite{Yunes}.  Our
analysis further supports the idea of a universal slope, dependent
only on the relevant critical collapse solution, but independent of the
initial conditions, and disfavours the astrophysical hypothesis of a
varying IMF.

Summarising, the dynamical exponents found in Newtonian and general
relativistic gravitational collapse can be used to model and
qualitatively explain a power law version of the IMF valid for small
masses.  For the first time, a concrete application to the numerical
phenomena of critical gravitational collapse has been proposed and
tested against observational data.  We have compared these results to
subsolar IMF data and found them in broad qualitative agreement for
low-mass systems, though quantitative agreement is poor at this stage
The key point is that gravitational collapse naturally leads to power
law behaviour in the low mass regime, with an exponent that is opposite
in sign to the observed high-mass behaviour.  This provides a new and
fresh view on power-law behaviour with specific astrophysical
applications to dynamic gravitational collapse and the IMF.

\section*{Acknowledgements}

We are grateful to Ben Owen and Tom Abel for insightful comments on
gravitational collapse theory and astrophysical observations of the
IMF.  Nicolas Yunes wishes to acknowledge the support of the Center
for Gravitational Physics and Geometry, the Center for Gravitational
Wave Physics and the Physics Department at The Pennsylvania State
University.

The research of Nicolas Yunes was supported partly by the US NSF
grants PHY-0114375 and PHY-0245649, while the research of Matt Visser
was supported by a Marsden Grant administered by the Royal Society of
New Zealand.



\end{document}